\documentclass[%
 11pt,
 amsmath,amssymb,twocolumn
]{article}

\usepackage{wrapfig}
\usepackage{authblk}
\usepackage{verbatim}
\usepackage[utf8]{inputenc}
\usepackage[english]{babel}
\usepackage{url}
\usepackage[margin=1in]{geometry}
\usepackage{graphicx}

\usepackage{hyperref}
\usepackage[dvipsnames]{xcolor} 
\hypersetup{
    colorlinks=true,
    citecolor=black,     
    urlcolor=RawSienna,
    linkcolor=BlueViolet,
}

\usepackage{dcolumn}
\usepackage{bm}
\usepackage{titlesec}
\usepackage{fancybox}
\usepackage[font=small,labelfont=bf]{caption}
\usepackage[belowskip=-15pt,aboveskip=0pt]{caption}

\usepackage[
    backend=biber,
    style=numeric, 
    citestyle=numeric,
    sorting=none, 
]{biblatex}
\titlespacing*{\section}
{0pt}{1ex}{1ex}
\titlespacing*{\subsection}
{0pt}{1ex}{1ex}
\titleformat*{\section}{\large\bfseries}
\titleformat*{\subsection}{\normalsize\bfseries}

\begin{document}

\renewcommand{\thefootnote}{\alph{footnote}}	

\begin{titlepage}
    \begin{center}
    \LARGE
    \textbf{Enabling and Enhancing Astrophysical Observations with Autonomous Systems} 
    \\ 
    \vspace{0.5cm}
    \normalsize
    Rashied Amini$^{1,}$\footnote{
    \href{mailto:rashied.amini@jpl.nasa.gov}{rashied.amini@jpl.nasa.gov}\\\textit{\copyright 2019 California Institute of Technology. Government sponsorship acknowledged.}}, Steve Chien$^{1}$, Lorraine Fesq$^{1}$, Jeremy Frank$^{^2}$, Ksenia Kolcio$^{3}$,\\
    Bertrand Mennsesson$^{1}$, Sara Seager$^{4}$, Rachel Street$^{5}$\\
    
    \vspace{0.25cm}
    July 10, 2019
    \vspace{0.15cm}
    
    \begin{figure}[h!]
        \centering
        \includegraphics[width=0.95\textwidth]{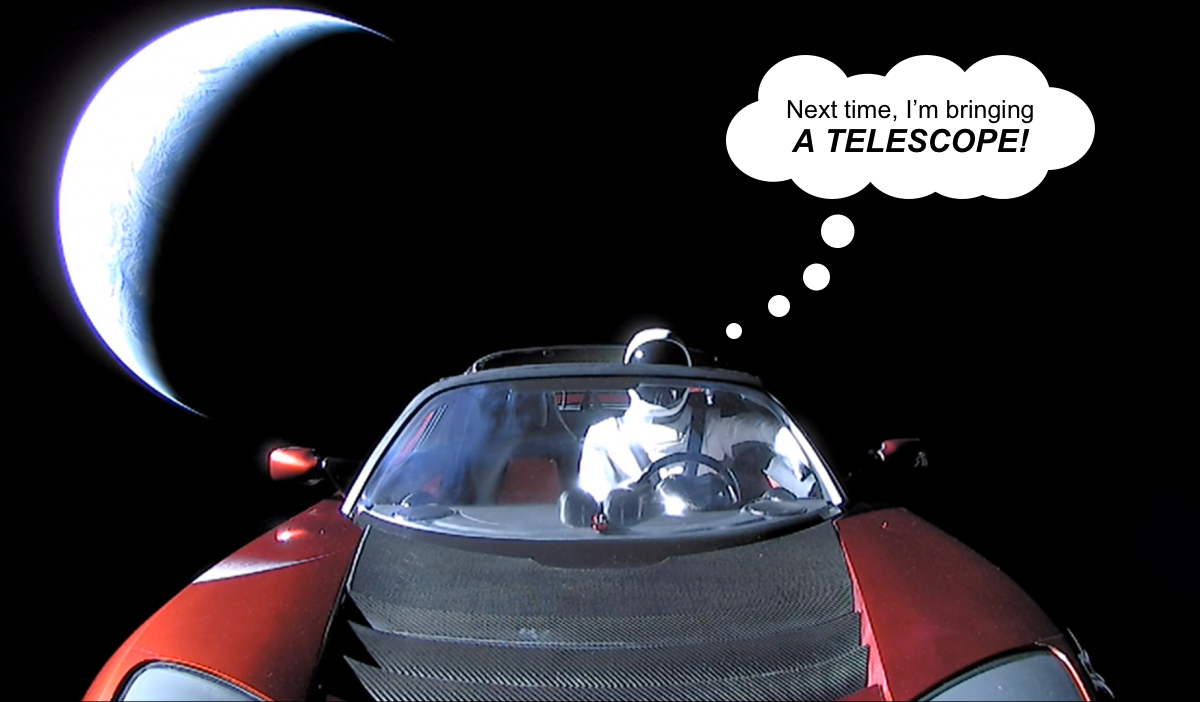}
    \end{figure}
    
    \vspace{0.5cm}
    \large
    \textbf{Endorsements}\\
    \normalsize
    Patricia Beauchamp$^{1}$, John Day$^{1}$, Russell Genet$^{6}$, Jason Glenn$^{7}$, Ryan Mackey$^{1}$, \\
    Marco Quadrelli$^{1}$, Rebecca Ringuette$^{8}$, Daniel Stern$^{1}$, Tiago Vaquero$^{1}$\\
    \vspace{1cm}  
    \normalsize
    $^{1}$\textit{NASA Jet Propulsion Laboratory}\\
    $^{2}$\textit{NASA Ames Research Center}\\
    $^{3}$\textit{Okean Solutions}\\
    $^{4}$\textit{Massachusetts Institute of Technology}\\
    $^{5}$\textit{Las Cumbres Observatory}\\
    $^{6}$\textit{California Polytechnic State University }\\
    $^{7}$\textit{University of Colorado at Boulder}\\
    $^{8}$\textit{University of Iowa}
    \end{center}
\end{titlepage}

\clearpage

\section{\label{sec:intro}Executive Summary}
Autonomy is the ability of a system to achieve goals while operating independently of external control \parencite{taxonomy}. The revolutionary advantages of autonomous systems are recognized in numerous markets, \textit{e.g.} automotive, aeronautics. Acknowledging the revolutionary impact of autonomous systems, demand is increasing from consumers and businesses alike and investments have grown year-over-year to meet demand. In self-driving cars alone, \$76B has been invested from 2014 to 2017 \parencite{brookings17}. In the previous Planetary Science Decadal, increased autonomy was identified as one of eight core multi-mission technologies required for future missions \parencite{board2012vision}.

The impact of autonomous systems on our ability to observe the universe can be just as revolutionary \cite{chien2017robotic}. However, relevant autonomy work to date has been limited in scope and too disjoint to confidently deliver anticipated capabilities, like in-space assembly (ISA), in a low risk and repeatable manner in the 2020s or even the 2030s. This paper includes the following so that the astrophysics community can realize the benefits of autonomous systems:
\begin{itemize}
    \item A description of autonomous systems with relevant examples
    \item Enabled and enhanced observations with autonomous systems
    \item Gaps in adopting autonomous systems
    \item Suggested recommendations for adoption by the Astro2020 Decadal
\end{itemize}

As we consider the observations necessary to answer new science questions formed in the 2010s, the need for autonomy is clear. Concept studies for the Astro2020 Decadal require operations that are more complex than ever before. Increasingly complex space- and ground-based observatories have more systems, components, and software. More engineering complexity invariably means that there are more paths for anomalies to disrupt a system's ability to perform its mission. This can reduce observational efficiency and potentially negate the advantages of larger apertures and more sensitive detectors. 

Servicing is a legal requirement for WFIRST and the Flagship mission of the 2030s \cite{s3729}, yet past and planned demonstrations may not provide sufficient future heritage to confidently meet this requirement. In-space assembly (ISA) is currently being evaluated to construct large aperture space telescopes \cite{ISA}. For both servicing and ISA, there are  questions about how nominal operations will be assured, the feasibility of teleoperation in deep space, and response to anomalies during robotic operation.

The past decade has seen a revolution in the access to space, with low cost launch vehicles, commercial off-the-shelf technology, and programs that have enabled numerous cubesat missions. NASA and academic institutions will be operating more small satellites and operations centers will need to adapt. The need will be greater if future human exploration goals to launch dozens of cubesats per SLS launch is met \cite{robinson2018nasa}. Operating autonomous observatories provides one solution to this impending problem. Notably, several ground-based observatories, like Las Cumbres and ALMA observatories, have begun using autonomous operations to command large arrays of telescopes, identifying advantages for observatories that follow their example. Planet and presumably SpaceX's Starlink, private space mission operators, have reached a break point where traditional commanding is inadequate to command their large constellations and are operating spacecraft with automated scheduling \cite{planet19}.

Gehrels/Swift is an inspiring example of the time-domain observations that autonomous systems enable. The multi-messenger approach for characterizing the physics leading to and resulting from gravity wave events will require similar missions to Gehrels/Swift. Gehrels/Swift relies on prescriptive state machines, statically-programmed conditions and routines also used in spacecraft fault protection, to execute autonomous Gamma-ray burst (GRB) follow-up observations. The system autonomy approach detailed in this paper offers several advantages over state machines in terms of dynamic decision-making and scalability. One major advantage is the ability to make decisions using on-board analysis of data to change an observation program. 

Dynamic decision-making also enables the restoration of functionality in the event of an anomaly. This type of decision-making is enabled by on-board health monitoring software, which monitors and diagnoses hardware anomalies to support autonomous systems. This results in greater observational efficiency and universally benefits all observatories. For observatories with competed time, this means more PIs can be supported. For mapping missions, like the Galaxy Evolution Probe, Probe of Inflation and Cosmic Origins, and Cosmic Dawn Intensity Mapper Probe, greater depths can be reached per unit time \cite{gep, pico, cdim}. For time-domain surveys, this results in less gaps in data.  

\begin{figure}[t]
    \centering
  \includegraphics[width=0.35\textwidth]{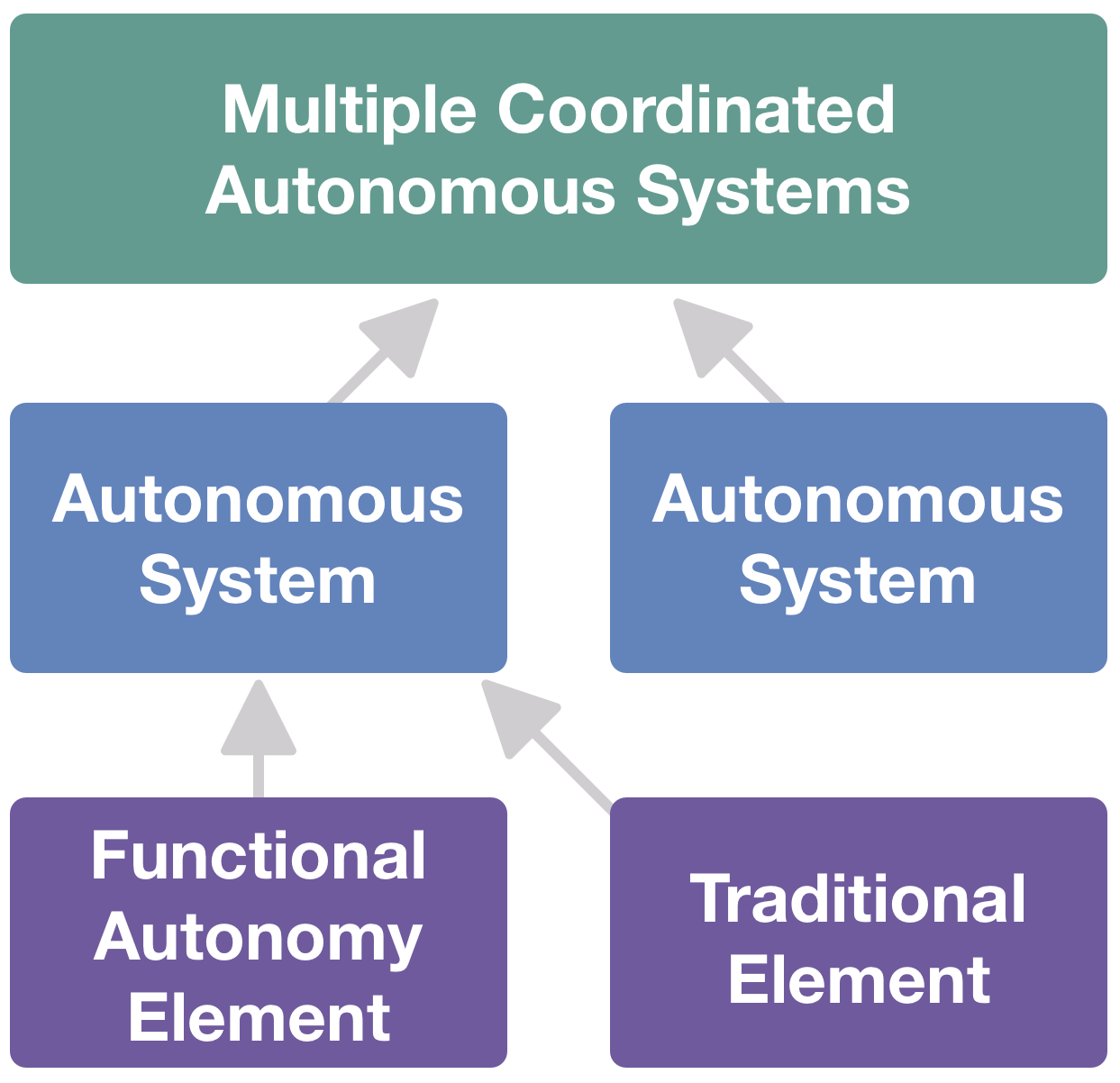}
  \caption{Effective, reliable autonomous systems must coordinate between the resources utilized by a system's lower level functions to achieve system-level goals. Appendix \ref{app-ex} offers an illustrated example of a system autonomy framework.}
  \label{fig:ha}
\end{figure}

As evidenced by private investments and developments in ground-based observatories, the adoption of autonomous systems in space is inevitable. There are two questions to the field: \textit{``When will we start using it?''} and \textit{``How will we start using it?''} Given the ambitions of the community, the time to begin is now. In order to use it in a repeatable, low risk, and cost-effective way, NASA, spacecraft vendors, and the astrophysics community need to cooperatively develop a coherent technical path forward. To do so, \textbf{our primary recommendation is for NASA to incentivize the use of autonomous systems for competed space missions,} for instance through a cost cap credit. Adoption in the 2020s will reduce the risk of future Flagship servicing missions.

\section{\label{sec:understandingAutonomy}Understanding Autonomous \\Systems}
Observing the proceedings of the Space Astrophysics Landscape in 2020 and Beyond meeting, it is clear that a gap exists between the expectations of the astrophysics community and the technical readiness of autonomy technologies required to meet these expectations. To understand this gap, we need to first define autonomy in a relevant context.

A hierarchy of systems is represented in Figure \ref{fig:ha}. At the bottom of the hierarchy is the functional-level, where control and autonomy is exercised in a limited domain. Functional control is the commanding actuators and sensors, \textit{e.g.} a command is sent and a motor turns at a commanded rate. Functional autonomy is decision-making within the boundaries of the functional element. A simple example is a state machine that (dis)engages a heater based on thermometer input. A more complicated example is an attitude controller that takes inputs of attitude knowledge (\textit{e.g.} star trackers). Its output is control system actuation to maintain a desired attitude. Pre-programmed routines filter inputs and evaluate conflicting knowledge, resulting in predictable behavior. 

More complex forms of functional autonomy have already been demonstrated and are currently being developed. For instance, autonomous optical navigation determines deviation from desired orbit ephemeris and has been used on Deep Space-1, Deep Impact/EPOXI, other planetary missions, and soon Arcsecond Space Telescope Enabling Research in Astrophysics (ASTERIA) \cite{autonav, asteriaLCPM}. On-going work on servicing and ISA utilizes computer vision as a knowledge source to control robotic actuation \cite{ISA}. On-orbit robotic servicing was first demonstrated on DARPA's OrbitalExpress in 2007 \cite{orbitalExpress}. In the next few years, RESTORE-L will be used to service Landsat-7 in low earth orbit using teleoperation after autonomous docking \cite{restore}.

However, functional elements utilize system resources, \textit{e.g.} time, power, attitude, data storage, \textit{etc}. Spacecraft are resource limited and efficient use is critical to mission success. Different activities may utilize resources in a mutually exclusive way; for instance, a space telescope may not be able to point its telescope at a target while simultaneously pointing its antenna toward Earth for communications. Some resources are zero-sum but accommodating of multiple spacecraft goals; for instance, all powered equipment require power but not all subsystem power modes can be supported simultaneously. Thus, there is a state of competition between different system goals. In the current state of practice, this competition is resolved by human planning during operations. Tools are used to define system activities, like observing and transmitting data, based on commands that are tied to certain resources. The goals of the scientists to observe the sky and goals of the engineers to preserve the spacecraft are merged using these tools to develop time-ordered sequences of commands that are uplinked to the spacecraft, \textit{e.g.} \cite{maldague1998apgen}. An extension of time-ordered sequences is conditional sequencing, where sequences use conditional statements as a state model. This approach has the capability of storing pre-defined routines that can later be executed \cite{grasso2008vml}. 

Autonomy poses a challenge to operational planning: how can you command a system that makes its own decisions? State machine-based autonomy is predictable in well-defined environments, and so resource budgets can be allotted because the input domain is well characterized. Spacecraft health is further ensured by fault protection state machines, adding another layer of protection. The use of state machines enables Gehrels/Swift to detect GRBs with the wide-field Burst Alert Telescope and slew to observe with its two other payloads \cite{swift}. However, machine learning-based decision-making and variable environments, exemplified by computer vision-guided robotic control, means that resource utilization cannot be effectively bounded in advance and so reliable, safe operation cannot be readily assured with traditional commanding. 

\begin{figure}[t]
      \centering
      \includegraphics[width=0.35\textwidth]{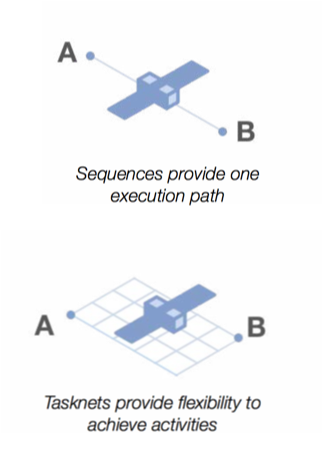}
      \caption{Task networks offer numerous pathways in time and state-space to achieve goals requested from ground operators. Implementing tasknet-based commanding enables ``push button-get science'' missions.}
      \label{fig:tasknets}
    \end{figure}

Coordination of resources used by functional elements, prescriptive or not, can be accomplished at the system-level through on-board planning and execution. This approach contrasts with traditional commanding with sequences through its use of task networks (tasknets, though goal and constraint networks are also used in the literature), described in Figure \ref{fig:tasknets}. Tasks are defined as commands that are associated with metadata defining the state conditions required for their execution and state impacts that result from their execution. Thus, graph networks of tasks can be constructed with tasks as nodes and edges connecting tasks whose state impacts are the state requirements of another task. Moreover, tasks can be have temporal constraints to be sequence-like. In this manner an autonomous system can be commanded like a traditional system if desired.

Sets of tasks can be defined as independent, uniquely prioritized system goals. Some goals identify system state transitions, such as the acquisition of new science data. Other goals identify states that need to be maintained and restored if lost, such as those related to spacecraft health. The role of on-board planning and execution is to negotiate between the constraints of all goals so that they can be executed without conflict or in violation of safe resources limits.

The final level of hierarchy at the top of Figure \ref{fig:ha} supports multiple autonomous systems in a multi-agent architecture.

\section{\label{sec:pastMissions}Examples of Relevant Autonomous Systems}
Spitzer Space Telescope, Dawn, Juno, and many other planetary missions have made use of conditional sequencing with the Virtual Machine Language (VML). Spitzer reported several advantages over traditional sequencing using VML. In particular, it made observations contingent on telescope settling state rather than sequenced time, which added one or two extra observations in an $11 \frac{1}{2}$ hour observing window. It also had the advantage of reducing spacecraft safing due to on-board memory overflow \cite{mittman2013scheduling}. However, as reported in \cite{mittman2013scheduling}, the limiting factor in implementing more of these autonomous behaviors was that there was ``no fast and effective way of modeling the flight system behavior on the ground.''

Autonomous systems relying on on-board planning and execution are beginning to see increased use on space- and ground-based observatories. A prominent example is the use of ASPEN/CASPER on Earth Observer-1, which used on-board science planning and execution to detect novel terrestrial scenes, like disasters, to autonomously perform follow-up observations \cite{aspen,tran2005autonomous}. Extending the work of CASPER, the Intelligent Payload EXperiment (IPEX) cubesat executed one year of autonomous payload operations using its on-board planner \cite{chien2014onboard, chien2017robotic}. PLan Execution Interchange Language (PLEXIL) is funded to be used to for a technology demonstration mission of multi-agent autonomy. Later in 2019, ASTERIA will demonstrate the use of the Multi-mission EXECcutive (MEXEC). Next year, Mars 2020 will use the Onboard Scheduler to maximize science return by using excess time and power at the end of each Martian sol to plan additional measurements \cite{fesq2019extended, verma2017autonomous, rabideau2017prototyping}. Temporal planning and scheduling systems also include IxTeT \cite{labone1995planning}, used for robotic contorl, and EUROPA \cite{barreiro2012europa}.  Other systems have been developed based on similar principles since then, notably IDEA and T-REX, used for autonomous underwater vehicles  \cite{mcgann2007t}. Appendix \ref{app-ex} offers an example of how these software are implemented in practice.

Las Cumbres Observatory and Atacama Large Millimeter/submillimeter Array (ALMA) are examples of ground-based observatories whose operations are autonomously planned and executed. Las Cumbres Observatory is a network of 18 telescopes at six sites that operate as a single observatory, enabling persistent observation. Scientist request observations, which are assigned and scheduled through a global scheduler \cite{volgenau2016two, boroson2014science}. General-purpose software has been developed for autonomous telescope operations that can be adopted by future observatories operating on these principles \cite{street2018general}. ALMA dynamically schedules and executes 30 minute ``scheduling blocks'' based on weather, science priority, project completion, and other parameters \cite{alma}.

Automated scheduling has traditionally been used for operational planning. Most relevantly, Space Telescope Institute uses SPIKE for planning Hubble observations \cite{johnston1990spike}. Planet uses automated scheduling to operate its fleet of earth observing cubesats. In human spaceflight Timeliner has seen significant use on-board the International Space Station (ISS) and is being considered as a candidate for the Lunar Gateway \cite{barnes2018autonomous}. While automated planning streamlines operations, it still has drawbacks when the plan cannot succeed due to operational conditions. 

\section{\label{sec:examples}Astrophysics with Autonomous Systems}

\subsection{Enabled Missions and New Science}
Autonomous systems have already enabled new astrophysics. Both ground- and space-based transient event observatories are fundamentally enabled by autonomous systems. Autonomous transient event detection and follow-up observation capability has been demonstrated with Gehrels/SWIFT and the Zwicky Transient Facility \cite{smith2014zwicky}. As exemplified by LCO, time-domain astronomy observations, \textit{e.g.} supernovae, microlensing, near earth asteroids, tidal disruption events, gravitational wave events, \textit{etc.} require real-time, highly reactive telescope scheduling. In these cases, observations cannot be planned in advance and the configuration of the observations may need to evolve over time according to the characteristics of event.  

With the projected improvements to ground-based detection and localization of gravity wave (GW) events, there is a need for observatories that can rapidly observe potential multi-messenger signals. Ground-based observatories will require the ability to respond to external signal, verify observability of the GW ellipse given current observatory conditions, and retask to perform GW follow-up observation while maintaining knowledge of the past observation. Space-based observatories will be required to do the same while also maintaining spacecraft health. Gehrels/SWIFT itself was launched in 2004 and may need replacement in the 2020s to retain the community's ability to perform GRB detection and localization over large areas of the sky. If ESA's Theseus is selected for M5, system autonomy software and expertise can serve as a potential NASA contribution to that mission.

Given the past priorities of the Astrophysics Decadal and NASA funding, it is expected that space-based time-domain observatories will be competed and are subjected to cost cap. For instance, the Gravitational-Wave Ultraviolet Counterpart Imager (GUCI) has already been proposed for the SmallSat call \cite{cenko2019gravitational}. These missions can be architected using state machine autonomy, following the Gehrels/SWIFT. Given the bounded nature of time-domain observations and the additional advantages afforded by on-board planning/execution, we note that these missions can alternatively use on-board planning/execution as a relatively low risk means of demonstrating and increasing the community's confidence in the technology.

As discussed in \cite{ISA}, system-level autonomy is required for ISA and servicing in order to coordinate robotic autonomy with the rest of the spacecraft. One example of how critical system-level autonomy is to ISA and servicing is the coordination of a mass model as robotic operation is performed. At a high-level, a servicer spacecraft has the goals of performing robotic operation and assuring attitude control in the presence of disturbance (gravity gradient, solar pressure). To accomplish the former goal, a robotic arm moves, changing the spacecraft's moment of inertia. To accomplish the second goal, the attitude control system maintains attitude based on a model of the spacecraft's moment of inertia. If robotic action is not coordinated, the attitude controller's moment of inertia model will not be consistent with reality. This may lead to over/under actuation of reaction wheels, potentially leading to collision risk and mission failure for both spacecraft.

Multi-agent autonomy also enables new observations. The AEON Network is ground-based facility currently under development operating numerous telescopes that will allow astronomers to submit requests for observation in real-time. Through multi-agent autonomy, a large network of ground- and space-based observatories, like AEON, can coordinate their observing programs across multiple facilities and wavelengths, serving as a powerful tool for characterizing new discoveries. Multi-agent autonomy can also be used on a constellations of low cost satellites as distributed transient event, namely GRB, observatories  \cite{racusin18,amini2008utilizing}. By using low cost scintillating detectors on low cost smallsat/cubesat platforms, localization can be performed through time-of-arrival similar to the Interplanetary Network. One advantage of this approach is the timeliness of observation. For instance, in simulations of flooding event observations by an earth observing constellation, a multi-agent architecture measured flood area to 96\% accuracy over time as opposed to a centrally planned architecture observing with 70\% accuracy, owing to the timeliness of observation \cite{nag}. Additionally, multi-agent coordination of more than two assets may be required, or would greatly facilitate, interferometry missions such as LISA. 

A unique class of missions that would benefit from ISA and multi-agent autonomy are radio and possibly NIR/optical/UV interferometry missions that may require ISA of large apertures and coordination. 

Autonomous systems also complement the increased access to space afforded by small satellites and low-cost launch vehicles. As the total number of missions increases, let alone missions that may utilize more than one spacecraft such as GUCI, the ground stations and operations facilities become a bottleneck for commanding and monitoring spacecraft. At some point, large numbers of traditional spacecraft cannot be efficiently commanded through traditional means. Autonomous systems reduce the human effort required to command as the burden can be off-loaded to an on-board planner.

Current plans for human exploration offer new platforms for astrophysics missions, creating new opportunities for the development of observatories. Most imminently, lunar exploration may create dozens of opportunities for new measurements. With cubesats piggybacking launches and opportunities to use the Lunar Gateway as a platform for payloads, managing multiple missions and scheduling observations that may have conflicting pointing and thermal requirements on Lunar Gateway becomes increasingly difficult to coordinate across multiple teams \cite{gateway}. Again, an operational bottleneck results that can be resolved through automated planning. Additionally, returning to the moon creates new opportunities for lunar surface-based observatories. This offers unique opportunities for some radio bands, cosmic ray, MeV $\gamma$-gay, X-ray, and UV measurements that cannot be made from Earth's surface. A Probe mission concept, FARSIDE, is a $\sim$10 MHz radio observatory on the farside of the Moon. As it requires a rover for deployment, autonomous mobility and robotic assembly capability is critical to mission feasibility. \cite{farside}

\subsection{Efficient Observing Programs}
The traditional paradigm of commanding reduces the overall efficiency of targeted observing programs as observation length is pre-determined in advance. Later, data is downlinked and analyzed on the ground. However, the efficiency of observing programs can be improved by analyzing data on-board to inform system decision-making.

One example is exoplanet direct imaging, exemplified by HabEx and LUVOIR, that requires a level of ~10$^{-10}$ raw contrast to perform direct imaging of exo-Earths. This raw contrast can only be effectively achieved in cases where exozodiacal light is not so bright that it reduces the effective raw contrast at the exoplanet's location. Even if exozodical light is previously characterized in mid-IR wavelengths \cite{mennesson2014constraining}, these observations may not predict the exozodiacal light at HabEx/LUVOIR near UV to near IR wavelengths. Additionally, not all systems will have constrained inclination that impacts the apparent brightness of the exozodiacal dust. Currently, HabEx and LUVOIR will schedule their observations in advance and use a pre-determined observing program based on little or no knowledge of the actual level of exozodi optical brightness around individual targets.  
 
In an autonomous system, coronagraphic imaging can be analyzed on-board the spacecraft to evaluate the contribution of exozodiacal light and the determine the value of continuing observation. In this case, excessive exozodical light can be detected on-board within a fraction of the planned observation time. On-board data processing software can then alert the on-board planner to truncate the observation so the observatory can perform the next scheduled observation. Data from the truncated observation is later downlinked for future analysis. In this example, more targets are observed more quickly, resulting in more observing time for other targets of interest and greater exo-Earth yield during the primary mission \cite{stark2015lower}.

\vspace{0.1cm}
\fbox{%
\begin{minipage}{0.45\textwidth} 
\textbf{Recommendation:} NASA should use ROSES as a means of funding software development for on-board data processing.
\end{minipage}}
\vspace{0.1cm}

The advantage of on-board data processing in union with a system planner is not limited to space-based observatories. Subsystems that evaluate weather and seeing conditions can aid to autonomously reschedule planned observations that may not be possible when scheduled, improving their net efficiency. 

\vspace{0.1cm}
\fbox{%
\begin{minipage}{0.45\textwidth}
\textbf{Recommendation:} NASA and NSF should incentivize the development of future ground-based observatories with automated scheduling/execution, following the example of ALMA and LCO.
\end{minipage}}
\vspace{0.1cm}

As discussed above, on-board data processing and multi-agent autonomy can be used in a coordinated network of ground- and earth-based to perform GW follow-up observatories. In such an architecture, localization that is currently performed \textit{post-hoc} as data is released can be performed on-board within the constellation, resulting in localization while the source is still emitting brightly.

\subsection{Adaptive Fault Protection Enables More Observations} 
Traditional space systems have fault protection schemes that enter safe modes, requiring human diagnosis and commanding to restore nominal operation. As a result, 4\% of nominal spaceflight operations are blocked by spacecraft safings \cite{imken2018modeling}. Notably, \cite{imken2018modeling} presents a lower bound on blocked operational time, as other anomalies can occur that restrict nominal operations and do not cause safing.

\begin{figure}[t]
  \centering
  \includegraphics[width=0.45\textwidth]{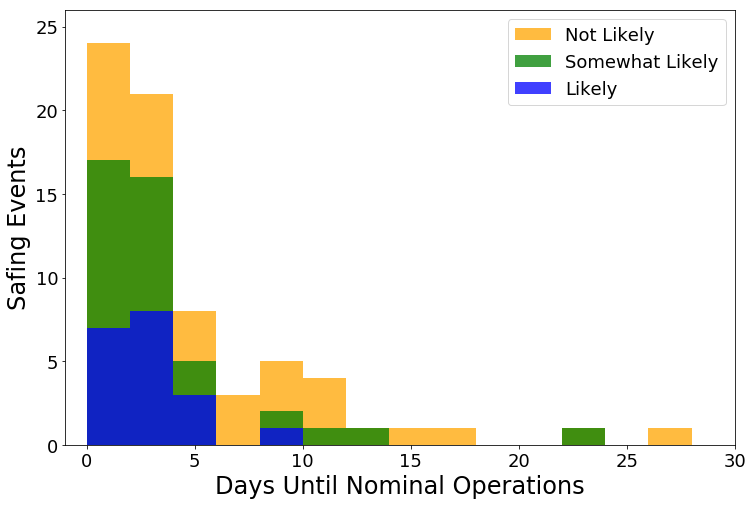}
  \caption{Histogram of safing events binned on the number of days between suspension and restoration of nominal operation. With on-board planning and execution and on-board health diagnosis, about 50\% of anomalies resulting safing may be averted. Result based on analysis of the \cite{imken2018modeling} safing dataset.}
  \label{fig:safings}
\end{figure}

Figure \ref{fig:safings} indicates that on-board planning/execution with on-board health diagnosis may mitigate the impact of about 50\% of safings. This adds an additional week of nominal operations per year. There are two reasons. Rather than relying on state machines for executing fault protection, health maintenance tasknets can restore the minimum functionality required to perform science operations while not endangering spacecraft health \cite{aaseng2018performance,frank2016transitioning}. Second, this architecture permits integration of on-board health diagnosis to monitor the health of hardware and local models for attitude knowledge and control, a major cause of safing events, to inform these health maintenance tasknets \cite{kolcio2014model,kolcio2016model}.

While an additional week of data per space observatory may seem marginal, if applied to NASA's fleet of space-based observatories it would result in 11 additional weeks of science per year for the community. The benefit is most useful to observatories with PI-directed observations, like Hubble and Spitzer, where additional PIs can be supported. For mapping missions, like GEP, additional mapping depth per unit time is achieved. For time-domain surveys, coverage is more complete in time. 

Even without on-board planning/execution, health diagnosis models and software can improve ground- and balloon-based observatories. Ballooning in particular suffers from a high failure rate, owing from \textit{ad hoc} integration of multiple payloads on-site and schedule constraints forcing limited testing. Recently, NASA JPL evaluated technologies for a self-reliant rover during which on-board health diagnosis was found to be effective in the build, integration, and testing environments in discovering and diagnosing hardware issues previously undetected \cite{gaines2016productivity,srr,kolcio2019model}. Health diagnosis software can be used for ballooning systems that are used repeatedly, such as mirror motor control, pressure vessels, and power generation, to detect hardware issues. This can reduce complexity and stress during the balloon integration phase and improve success rate of balloon missions.

\vspace{0.2cm}
\fbox{%
\begin{minipage}{0.45\textwidth}
\textbf{Recommendation:} Integrate the use of health diagnosis software for elements that are repeatedly used on ballooning platforms.
\end{minipage}}
\vspace{0.2cm}

\section{Addressing the System Autonomy Gap}
For astrophysics, autonomous systems can enable and enhance missions that deliver revolutionary data sets, reduce the cost of missions, and reduce the burden on scientists in developing and maintaining observing programs. A future where ``press button $-$ get science'' missions is on the horizon, but work remains that requires the community's awareness and support.

The primary gap is cultural. Autonomous systems imply a different paradigm of design and operations compared to traditionally-commanded systems. Compared to autonomy in the private sector, a small proportion of our space science and spaceflight communities have relevant expertise to review the opportunities and risks associated with autonomous systems. This is compounded by traditional engineering preference for heritage designs and expectations of predictability. On point, how can scientists, engineers, and proposal reviewers be confident in a mission concept that operates itself? Is it possible to design and deploy autonomous systems that are partially autonomous to placate the concerns of the community? These questions need to be formally addressed if NASA is to meet its legal requirement to perform servicing requirement for future large, space-based observatories, let alone to reap the benefits of autonomous systems for observation.

Limited institutional capacity to adopt autonomous systems is exemplified by the examples of autonomous systems above: most of these missions were or will be designed and built by NASA. Given the high cost and risk associated with changing the process by which spacecraft are designed, built, and tested, spacecraft vendors have till now relied on conditional sequencing and not autonomous planning/execution. Thus, government-industry cooperation is required to make use of autonomous systems reliably and repeatably for all NASA missions.

\vspace{0.2cm}

\fbox{%
\begin{minipage}{0.45\textwidth}
\textbf{Recommendation:} NASA should incentivize the use of autonomous systems for competed space missions. Specifically, small sat missions, missions of opportunity, SMEX, MIDEX, and Probe missions can include a credit for using the technology. \textit{We note that transient event observatories offer a low risk path to maturing this critical technology.}
\end{minipage}}\\

\vspace{0.1cm}

Another aspect of the cultural gap is NASA's definition of technology readiness and its reference to ``operational environment'' that is overly restrictive to software technologies that can be effectively validated outside of the operational environment, \textit{e.g.} on-board science data processing software.

\vspace{0.1cm}

\fbox{%
\begin{minipage}{0.45\textwidth}
\textbf{Recommendation:} NASA should evaluate the applicability of the Technology Readiness Level as a means of evaluating the maturity of autonomy and on-board data processing software. 
\end{minipage}}
\vspace{0.1cm}

Other gaps are technical. Autonomy frameworks, described in Appendix \ref{app-ex}, define rules for how system-level planning and execution interface with traditional components and systems and functional autonomy. Community acceptance of these frameworks can reduce adoption risk and promote repeatability by permitting traditional design and operations approaches. By defining a convention for how autonomous missions should be designed and built, frameworks would also improve reviewability of autonomous missions and portability of testing methodology. Remaining work includes improving the verifiability of tasknets, which is critical to reaping the benefits of integrated fault protection. Relatedly, telemetry that permits reconstruction of on-board decision-making requires further study and definition. Ground systems and tools for commanding of autonomous spacecraft require further maturation.

Finally, some observations will benefit from on-board data analysis. For these observations, new software will be required to perform this function, which will be the responsibility of science community. While not the subject of this white paper, processing-intensive data processing may require high performance computing. High performance computing does not necessarily enable autonomous systems, but is enhancing by permitting intensive processing of science data and on-board scheduling over larger search spaces.

\vspace{0.2cm}
\fbox{%
\begin{minipage}{0.45\textwidth}
\textbf{Recommendation:} NASA should fund technology demonstrations of high-performance space-based computing for on-board data processing.
\end{minipage}}
\vspace{0.2cm}

\appendix
\section{Example of a System Autonomy Framework}
\label{app-ex}
\begin{figure*}[h!]
  \centering
  \includegraphics[width=0.75\textwidth]{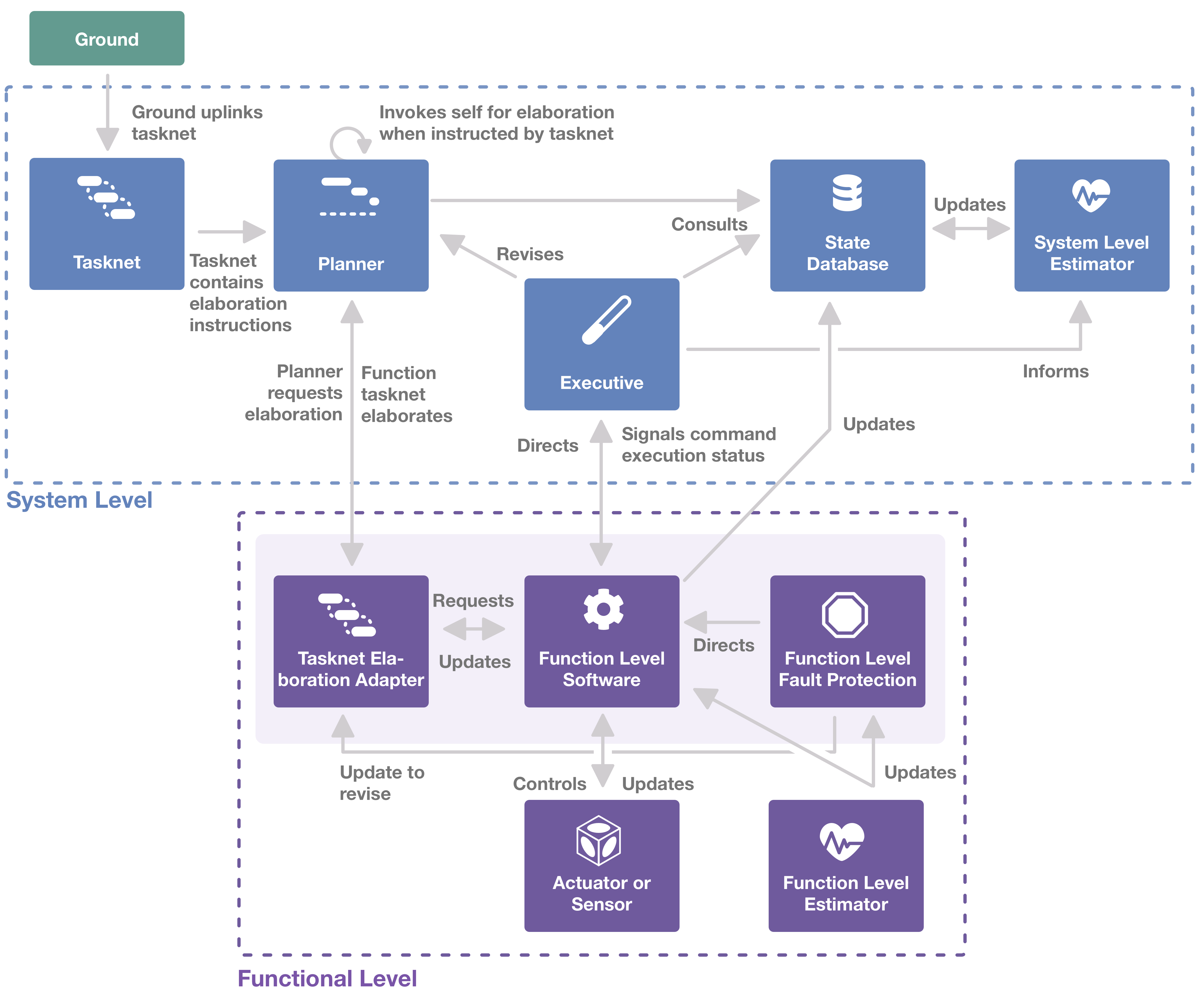}
  \caption{An example of an autonomous system framework, the Framework for Robust Execution and Scheduling of Commands On-Board (FRESCO), defines capabilities and interfaces that will resulting in repeatable and predictable implementations of systems autonomy for complex systems, such as spacecraft. This figure offers a simplified description of FRESCO components and interfaces.}
  \label{fig:fresco}
\end{figure*}

In order to create autonomous systems repeatably and reliably, a framework must be defined. Similar to a legal constitution, a system autonomy framework defines responsibilities and capabilities of system components and how they interface with one another in governing system behavior. For instance, the manner in which on-board science data processing is interfaced to inform decisions at a system-level should be identical across all missions regardless of which decisions it informs. This enables reliable mutli-mission adoption of on-board planning/execution. Under a unified framework, engineers, scientists, and managers can work toward the same set of requirements that assure mission success. Reviewers can also use the same framework to verify compliance. Without a unified framework, development, mission assurance, and review can become intractable given the complexity in designing autonomous systems. 

There are several requirement that define an effective autonomy framework. It should make guarantees about acceptable behavior, enable confident operator oversight and insight, readily accommodate new information, and not require extensive tailoring or \textit{ad hoc} modification to support multiple missions. Pragmatically, such a framework must afford a practical path toward adoption. To do so, system autonomy must integrate with existing components. Human workflows involved across mission phases should deviate minimally from existing practice. Also, the framework must support varying degrees of autonomy $–$ permitting sequence-like commanding to highly autonomous operation. Without this practical path, NASA and industry partners will have to invest in brand new software and processes and accept significant risk in implementing a major leap toward systems autonomy at once.

There are several examples of such a framework. The Framework for Robust Execution and Scheduling of Commands On-Board (FRESCO) is under development at NASA JPL; the NASA Platform for Autonomous Systems (NPAS) is under development at NASA SSC; and, a vehicle management system for autonomous spacecraft habitat operations is under development at NASA ARC and JSFC \cite{asteriaLCPM,figueroa2019nasa,levinson2018development}. Finally, the European Robotic Goal-Oriented Autonomous Controller (ERGO) has been under developed by an EU-funded consortium of industry and academia \cite{ocon2018ergo}. Below, we use FRESCO as an example to illustrate how systems autonomy is implemented.

\textbf{Tasknet} Tasknets are data structures that encapsulate the potential envelop of spacecraft behavior. They are graph networks where nodes are tasks and the edges are the state and temporal dependencies between tasks. Tasknets can be defined as goals for the spacecraft to achieve, to transition states (\textit{e.g.} an imaging survey goal results in a set of images being taken) and to maintain states (\textit{e.g.} a pointing knowledge maintenance goal restores pointing knowledge through optical navigation if a knowledge uncertainty threshold is violated). Tasknets have been described in literature since the 1970s \cite{strips,vere1985deviser}. 

\textbf{Planner and Executive} Planners create and maintain schedules of tasknets based on their prioritization and projected timelines of future states. Scheduling tasks is performed by a search function, whose search space can be constrained based on how tasks are defined. This permits traditional, sequence-like behavior or highly autonomous behavior within the same framework.  At a certain time before scheduled execution, the planner passes tasks to the executive for execution. Executives are responsible for intelligent execution and monitoring the impact of executed tasks. Under nominal operation, they receive receipt of successful task execution and proceed to dispatch the next scheduled tasks. If a task fails, they can exercise contingency behaviors specified by the task, which can include replanning requests to the planner.

Currently, MEXEC and PLEXIL  are maintained by NASA JPL and ARC, respectively \cite{verma2017autonomous, verma2006universal}. A wider survey of command execution systems is presented in \cite{verma2005survey}.

\textbf{State Database} A state database serves as a ``single source of truth'' for the system, maintaining component status and abstracted system states used in decision-making.

\textbf{System-Level Estimator} Estimators that inform decision-making use system telemetry as input to models of system behavior. System health monitoring software is one such estimator. System health monitoring serves two purposes. First, it can be used to identify potentially faulty components to alert operators to potential future risks. In rover testing, it was able to successfully identify undiagnosed hardware problems\cite{kolcio2019model}. Second, it permits the creation of tasknets that operate only if healthy component states are reported, reducing the risks of autonomous operation. MONSID, developed by Okean Solutions, uses linked models of hardware behavior to monitor the health status of components \cite{kolcio2016model,kolcio2017model,kolcio2019model}.

\textbf{Function-Level Software and Components} Function-level software can perform a multitude of functions, ranging from hardware control to data processing and functional autonomy. Depending on its function, its internal interfaces will vary. For instance, if a hardware controller includes local fault protection, an interface for a signal to interrupt system-level execution over that controller's domain is required. Traditional hardware controllers and on-board data processing can also be used to inform the scheduling of tasknets. The on-board processing for exozodical light in exoplanet coronagagraphy serves as an example. 

\section{Acknowledgements}
Thank you to Ellen Van Wyk, NASA JPL, for illustrations included in this white paper and SpaceX for the cover photograph.

\clearpage
\printbibliography
\end{document}